\documentclass[aps,prl,a4paper,preprintnumbers,tightenlines,superscriptaddress,
twocolumn,floatfix,showpacs,final,nofootinbib]{revtex4-1}
\usepackage{graphicx}
\usepackage{longtable}
\usepackage{amsmath}
\usepackage{amssymb}
\usepackage{subfigure}
\usepackage{hyperref}
\usepackage{dcolumn}
\usepackage{bm}
\usepackage{color}

\newcommand{\dd}{\, {\rm d}}
\newcommand{\gsim}{\;\mbox{\raisebox{-0.5ex}{$\stackrel{>}{\scriptstyle{\sim}}$}
}\;}
\newcommand{\lsim}{\;\mbox{\raisebox{-0.5ex}{$\stackrel{<}{\scriptstyle{\sim}}$}
}\;}

\newcommand{\kar}{\kappa_{-2}}

\newcommand*\colvec[3][]{\begin{pmatrix}\ifx\relax#1\relax\else#1\\\fi#2\\#3\end{pmatrix}}

\newcommand{\lhb}{L_{\rm HB}}

\newcommand{\ccc}{{_{\rm c}}}

\newcommand{\eee}{_{\rm e}}

\def\eea{\end{eqnarray}}
\def\bea{\begin{eqnarray}}
\allowdisplaybreaks
\usepackage{enumitem}
\setenumerate[1]{label=(\arabic*)}

\begin{document}
\title{Hydrogen Burning in Low Mass Stars Constrains Alternative Gravity Theories}
\author{Jeremy Sakstein}
\email[Email:]{jeremy.sakstein@port.ac.uk}
\affiliation{Institute of Cosmology and Gravitation,
University of Portsmouth, Portsmouth PO1 3FX, UK}

\begin{abstract}

The most general scalar-tensor theories of gravity predict a weakening of the gravitational force inside astrophysical bodies. There is 
a minimum mass for hydrogen burning in stars that is set by the interplay of plasma physics and the theory of gravity. We calculate this for 
alternative theories of gravity, and find that it is always significantly larger than the general relativity prediction. The observation of several 
low mass Red Dwarf stars therefore rules out a large class of scalar-tensor gravity theories, and places strong constraints on the cosmological 
parameters appearing in the effective field theory of dark energy.

\end{abstract}
\maketitle

The dark energy and cosmological constant problems have been driving the study of alternative theories of gravity as a possible explanation for the 
acceleration of the cosmic expansion (see \cite{Joyce:2014kja} for a recent review). Among them, scalar-tensor theories are particularly well-studied, 
which raises the question of what is the most general theory that is free of the Ostrogradsky ghost instability. This has recently been answered 
by Gleyzes, Langlois, Piazza \& Vernizzi (GLPV) \cite{Gleyzes:2014dya}. 
The GLPV framework encapsulates every healthy scalar-tensor theory, and a large sub-class, the \textit{beyond Horndeski class}, admit 
self-accelerating solutions that are viable competitors to $\Lambda$CDM \cite{Kase:2014yya,Barreira:2014jha}. These theories have a peculiar 
property: they make identical predictions to general relativity (GR) except that they predict a weakening of the strength of gravity inside 
astrophysical bodies \cite{Kobayashi:2014ida}. 
% 
% 
% % {\bf Every healthy scalar-tensor theory is encapsulated 
% % within this framework, including well-studied models such as Brans-Dickie, $f(R)$, quintessence, k-essence, DBI, chameleon, symmetron and disformal 
% % models, as well as self-accelerating theories such as galileons. They are the starting point for any new theory of scalar-tensor gravity and thus any 
% % model-independent constraint has the potential to drastically reduce the viable theory space. 
% 
% GLPV} theories 
% 
% 
% 
% 
This has an important implication which has 
hitherto been left unexplored: the minimum mass for hydrogen burning (MMHB) is larger than the observed mass of several hydrogen burning Red 
Dwarf stars. Low mass stellar objects do not achieve the necessary core conditions to ignite hydrogen, but heavier objects are hotter and denser. 
Stars heavier than the MMHB fuse hydrogen and are classified as Red Dwarfs; those with lower masses are classified as brown dwarfs. General 
relativity predicts that the MMHB is $0.08M_\odot$ \cite{1963ApJ...137.1121K,Burrows:1992fg} and, indeed, several Red Dwarfs have been observed with 
masses larger than $0.08M_\odot$ (see \cite{Chabrier:2008bc} and references therein). In this paper we calculate the MMHB predicted by the most 
general scalar-tensor theories. The reduced strength of gravity has the effect that hydrostatic equilibrium can be maintained at lower core 
temperatures and densities. As a consequence, we find that the MMHB can be significantly larger than observed in nature, and therefore a large 
class of these models can be ruled out. A large majority of current dark energy research focuses on models that are subsets of the GLPV theory 
\cite{DeFelice:2011hq}. Therefore, our results have important implications for scalar-tensor dark energy models. This is the first time 
that the observation of hydrogen burning in stars has been used to constrain the theory of gravity.

The structure of low mass stellar objects makes them perfect probes of modified gravity. They have uniform compositions and have properties that are 
only weakly dependent on non-gravitational physics such as variations in their metallicity and opacity. This means they are free of the degeneracies 
that plague some \cite{Davis:2011qf,Sakstein:2013pda,Vikram:2014uza}, but not all, \cite{Jain:2012tn} stellar structure tests of gravity. 
Furthermore, their equation of state is 
well-known, and is relatively simple. This means that the polytropic techniques developed by \cite{Koyama:2015oma,Saito:2015fza} can be applied to 
calculate the 
MMHB analytically. \cite{Koyama:2015oma,Saito:2015fza} have shown that scalar-tensor theories predict that hydrostatic equilibrium equation is 
modified to
\begin{equation}\label{eq:MGHSE}
\frac{\dd P}{\dd r}=-\frac{GM\rho}{r^2}-\frac{\Upsilon}{4}G\rho\frac{\dd^2 M}{\dd r^2}.
\end{equation}
$\Upsilon$ is the only free dimensionless parameter that characterises the deviations from GR. When $\Upsilon>0$, the new term is 
negative because the mass is more concentrated in the centre of the star, therefore the strength of gravity is reduced compared with GR. 
%
%{\bf The conservation of mass implies $\dd M/\dd r=4\pi 
%r^2\rho(r)$, and differentiating, one finds $\dd ^2M/\dd r^2$ scales like $\dd \rho/\dd r$. The stellar density decreases outwards and so the second term in (\ref{eq:MGHSE}) 
%is negative.}
%
%
%Note that the conservation of mass implies $\dd M/\dd r=4\pi 
%r^2\rho(r)$ and so $\dd ^2M/\dd r^2$ scales like $\dd \rho/\dd r$. The stellar density decreases outwards and so the second term in (\ref{eq:MGHSE}) 
%is negative. 
%This shows how the strength of gravity is reduced in scalar-tensor theories. 
%In terms of the fundamental theory, $\Upsilon$ is a 
%combination of the new parameters appearing the Lagrangian and the time-derivative of the cosmological scalar \cite{Kobayashi:2014ida}. One example of a well-studied and viable alternative to GR is the covariant quartic galileon \cite{Kase:2014yya}, which 
%admits self-accelerating solutions without the need for a cosmological constant. 

The parameter $\Upsilon$ is directly related to the parameters $\alpha_B$, $\alpha_T$ and $\alpha_H$ appearing in the effective field theory 
(EFT) of dark energy \cite{Gleyzes:2014qga} via \cite{Saito:2015fza}
\begin{equation}
\Upsilon = \frac{4\alpha_H^2}{\alpha_H-\alpha_T-\alpha_B(1+\alpha_T)}.
\end{equation}
The five parameters appearing in the EFT completely characterise the linear cosmology of the theory\footnote{Note that there are two parameters, 
$\alpha_K$ and $M$, that are not constrained by the effects presented here.}. Any independent constraint on $\Upsilon$ therefore constrains this 
combination of the cosmological parameters, complimenting searches on other scales and restricting the possible deviations from GR. One example of a 
well-studied and viable alternative to GR is the covariant quartic galileon \cite{Kase:2014yya}, which 
admits self-accelerating solutions without the need for a cosmological constant. This theory has $\Upsilon=1/3$, which, as we will see below, is 
ruled out by the constraint we will ultimately obtain using the MMHB.
 
Stars are the result of the complex interplay between many 
different areas of physics, but this is the only stellar structure equation where the modifications of 
GR appear. The rest of the physics that determines the structure and evolution of the star---energy generation from nuclear burning, 
the equation of state, the opacity and composition---is non-gravitational, and is hence unaltered by changing the theory of gravity. 

Low-mass stellar objects\footnote{See \cite{Burrows:1992fg} 
for a review of the science of brown dwarfs and other low-mass stellar objects.} are supported by a combination of thermal and electron degeneracy 
pressure, and are well-described by the polytropic equation of state \cite{Burrows:1992fg}
\begin{equation}\label{eq:Kdef}
 P=K\rho^{\frac{5}{3}},\quad K=\frac{(3\pi^2)^{\frac{2}{3}}\hbar^2}{5m\eee m_{\rm 
H}^{\frac{5}{3}}\mu\eee^{\frac{5}{3}}}\left(1+\frac{\alpha}{\eta}\right),
\end{equation}
where $m\eee$ and $m_{\rm H}$ are the electron and hydrogen mass respectively, $\mu\eee=1.143$, and $\alpha=4.82$. $\eta$ is a measure of the 
degeneracy of the star. It is formally defined as the ratio of the Fermi energy to $k_{\rm B}T$, and is given by
\begin{equation}
\eta =\frac{(3\pi^2)^{\frac{2}{3}}\hbar^2}{2m_{\rm e}m_{\rm H}^{\frac{2}{3}}k_{\rm B}}\frac{\rho^{\frac{2}{3}}}{\mu_{\rm e}^{\frac{2}{3}}T}.
\end{equation}
$\eta$ is a constant throughout the entire star. Defining the dimensionless radial coordinate and density variables
\begin{equation}\label{eq:polydef}
 r=r\ccc\xi,\quad r\ccc^2=\frac{5K}{8\pi G}\rho\ccc^{2},\quad \rho=\rho\ccc\theta^{\frac{3}{2}},
\end{equation}
where $\rho\ccc$ is the central density, the structure of the star is governed by the Lane-Emden equation (LEE) \cite{Koyama:2015oma}
\begin{equation}\label{eq:MLE}
 \frac{1}{\xi^2}\frac{\dd}{\dd \xi}\left[\left(1+\frac{3{\Upsilon\xi^2\theta^{\frac{1}{2}}}}{8}\right)\xi^2\frac{\dd \theta}{\dd 
\xi}+\frac{\Upsilon}{2}\xi^3\theta^{\frac{3}{2}}\right]=-\theta^{\frac{3}{2}}.
\end{equation}
The boundary conditions are $\theta(0)=1$ and $\theta'(0)=0$, which imply that, near the origin
\begin{equation}\label{eq:thexpo}
 \theta(\xi)\approx 1-\frac{1}{6}(1+\frac{3\Upsilon}{2})\approx \exp\left[-\frac{1}{6}(1+\frac{3\Upsilon}{2})\right].
\end{equation}
The radius of the star, $R$, is defined by $\rho(R)=0$, which defines $\xi_R$ such that $\theta(\xi_R)=0$. The structure of the star is described by 
solutions of the LEE, which predicts the stellar radius, the mass-radius relation and the central density:
\begin{align}\label{eq:poly2}
M&=4\pi r\ccc^3\rho\ccc\omega,\,\, R=\gamma\frac{K}{GM^{\frac{1}{3}}},\,\,\rho\ccc=\delta\frac{3M}{4\pi R^3},
\end{align}
where the ($\Upsilon$-dependent) structure coefficients are
\begin{align}
\omega&\equiv 
-\xi_R^2\left.\frac{\dd\theta}{\dd\xi}\right\vert_{\xi={\xi_R}},\label{eq:ome}\quad\gamma\equiv\left[\frac{125}{128\pi^2}\right]^\frac
{ 1 } { 3 } \omega^{\frac{1}{3}}\xi_R\\& \textrm{and}\quad\delta\equiv-\frac{\xi_R}{3\dd\theta/\dd\xi|_{\xi=\xi_R}}.
\end{align}
These quantities then encode the effects of scalar-tensor theories on the stellar properties.

The rate-limiting reaction for hydrogen burning is the weak process $\textrm{p}+\textrm{p}\rightarrow\textrm{d}+\textrm{e}^++\nu_e$. The energy 
generation rate (per unit mass) at typical brown dwarf temperatures and densities can be approximated by the power-law \cite{Burrows:1992fg}
\begin{equation}\label{eq:engen}
\epsilon_{\rm HB}=\epsilon\ccc\left(\frac{T}{T\ccc}\right)^s\left(\frac{\rho}{\rho\ccc}\right)^{u-1},\quad \epsilon_c=\epsilon_0 
T\ccc^s\rho\ccc^{u-1},
\end{equation}
with $s\approx 6.31$, $u\approx 2.28$, and $\epsilon_0=3.4\times10^{-9}$ ergs/g/s. Our procedure for calculating the MMHB will then be the following: 
We will first calculate the luminosity in hydrogen burning, $\lhb$, by integrating 
(\ref{eq:engen}) over the entire star using solutions of the LEE. 
Next, the 
luminosity at the photosphere $L\eee$ is calculated using the Stefan-Boltzmann law. The MMHB is the lowest mass where the reaction above can be 
sustained stably, i.e. energy generated in the core is compensated by energy radiated from the surface, which corresponds to the mass where 
$\lhb=L\eee$. Demanding that the two are equal then gives us a condition for the mass.

We begin with the luminosity in hydrogen burning. Since $\eta$ is a constant, we have $T/T\ccc=(\rho/\rho\ccc)^{\frac{2}{3}}$ and so, using 
Eqns. (\ref{eq:polydef}) and (\ref{eq:engen}) we have
\begin{equation}\label{eq:lnint1}
\lhb=4\pi r\ccc^3\rho\ccc\epsilon\ccc\int_0^{\xi_R}\xi^2\theta^{\frac{3}{2}u+s}\dd\xi.
\end{equation}
This can be integrated using Eqns. (\ref{eq:thexpo}) and (\ref{eq:poly2}) to find
\begin{equation}
\lhb=\frac{3\sqrt{\pi}}{\sqrt{2}\omega\left[(1+\frac{3\Upsilon}{2})(\frac{3}{2}u+s)\right]^{\frac{3}{2}}}\epsilon\ccc M.
\end{equation}
Using Eqns. (\ref{eq:Kdef}) and (\ref{eq:poly2}), one has
\begin{align}
 \rho\ccc&=\frac{125 G^3 m\eee^3  m_{\rm H}^5\mu\eee^5}{12\pi^5\hbar^6}\frac{\delta}{\gamma^3} 
M^2\left(1+\frac{\alpha}{\eta}\right)^{-3}\label{eq:rhocfull}\quad\textrm{and}\\
T\ccc&=\frac{25 G^2 m\eee m_{\rm H}^{\frac{8}{3}} \mu\eee^{\frac{8}{3}}}{2^{\frac{7}{3}}\pi^2 k_{\rm 
B}\hbar^2}\frac{M^{\frac{4}{3}}}{\gamma^2}\frac{\eta}{(\alpha+\eta)^2}\label{eq:Tcfull},
\end{align}
which can be used in equation (\ref{eq:engen}) to find $\epsilon\ccc$. The luminosity in hydrogen burning is then
\begin{align}\label{eq:LHB}
\lhb= {{5.2\times10^6} L_\odot}
\frac{\delta^{5.487}}{\omega\gamma^{16.46}(1+\frac{3\Upsilon}{2})^{\frac{3}{2}}}M_{-1}^{11.973}\frac{\eta^{10.15}}{
(\eta+\alpha)^{16.46}},
\end{align}
where $M_{-1}=M/0.1M_\odot$.

Our next task is to compute the luminosity at the photosphere. Using numerical modelling of the equation of state of hydrogen at high densities, it 
has been shown that the temperature and density at the photosphere are related via \cite{Burrows:1992fg}:
\begin{equation}\label{eq:Trhoana}
\frac{T\eee}{{\textrm{K}}}=\frac{1.8\times10^6}{\eta^{1.545}}\left( \frac{\rho\eee}{\textrm{g/cm}^3}\right)^{0.42}\,.
\end{equation}
The photosphere is very thin compared with the radius of the star, and so the surface gravity, $g=GM(r)/r^2$, can be 
treated as a constant to a high degree of accuracy. The location of the photosphere is the radius where the optical depth, defined as 
\begin{equation}\label{eq:optdepth}
 \tau(r)=\int_{r\eee}^\infty \kappa \rho\dd r,
\end{equation}
where $\kappa$ is the Rosseland mean opacity, is equal to $2/3$. In GR, this is found by substituting for the density using the hydrostatic 
equilibrium 
equation (\ref{eq:MGHSE}) and integrating using the fact that $g$ is constant. In our theory, there is the additional complication coming from the 
term proportional to $\dd ^2M/\dd r^2$. We will deal with this using the same approximation as is made in the GR calculation i.e. assuming that $g$ 
is constant. This implies that $\dd(M/r^2)/\dd r=0$ and hence we have 
\begin{equation}
 \frac{\dd M}{\dd r}=2\frac{M}{r}\quad\textrm{and}\quad \frac{\dd ^2 M}{\dd r^2}=2\frac{M}{r^2}.
\end{equation}
The hydrostatic equilibrium in the vicinity of the photosphere is then 
\begin{equation}
 \frac{\dd P\eee}{\dd r}=-g\rho\left(1+\frac{\Upsilon}{4}\right).
\end{equation}
Substituting this into (\ref{eq:optdepth}) we find
\begin{equation}\label{eq:Pphot}
  P\eee=\frac{2}{3\kappa}\left(1+\frac{\Upsilon}{2}\right)g=\frac{\rho\eee k_{\rm B} T\eee}{\mu m_{\rm H}},
\end{equation}
where the ideal gas law has been used and the mean molecular mass $\mu=0.593$ corresponding to a hydrogen-helium mixture. Using Eqn. 
(\ref{eq:poly2}), the surface gravity is
\begin{align}
g=3.15\times10^{6}\gamma^{-2}M_{-1}^{\frac{5}{3}}\left(1+\frac{ \alpha}{\eta}\right)^{-2}\textrm{cm}/\textrm{s}^2.
\end{align}
Using this in Eqn. (\ref{eq:Pphot}) and replacing the pressure using the Polytropic relation (\ref{eq:polydef}), we find the density at the 
photosphere
\begin{align}\label{eq:rhocana}
\frac{\rho\eee}{\textrm{g}/\textrm{cm}^3}&=5\times10^{-5}M_{-1}^{1.17} 
\left[\frac{(1+\frac{\Upsilon}{2})}{\kar}\right]^{0.7}\frac{\eta^{1.09}}{\gamma^{1.41}}\left(1+\frac{\alpha}{\eta
}\right)^{-1.41},
\end{align}
where $\kar=\kappa/10^{-2}\,\,\textrm{cm}^2/\textrm{g}$. This can then be used in conjunction with Eqn. (\ref{eq:Trhoana}) to find 
the effective temperature
\begin{equation}
\frac{T\eee}{\textrm{K}}=2.9\times10^4\frac{ M_{-1}^{0.49} }{\gamma^{0.59}\eta^{1.09}}
\left[\frac{(1+\frac{\Upsilon}{2})}{\kar}\right]^{0.296}\left(1+\frac{\alpha}{\eta}
\right)^{-0.59}.
\end{equation}
Using these in the formula $L\eee=4\pi R^2\sigma 
T\eee^4$ we find
\begin{equation}\label{eq:L}
L\eee=2.65L_\odot 
\frac{M_{-1}^{1.305}}{\gamma^{2.366}\eta^{4.351}}\left[\frac{(1+\frac{\Upsilon}{2})}{\kar}\right]^{1.183}\left(1+\frac{\alpha}{\eta}\right)^{-0.366}.
\end{equation} 

\begin{figure}
\includegraphics[width=0.45\textwidth]{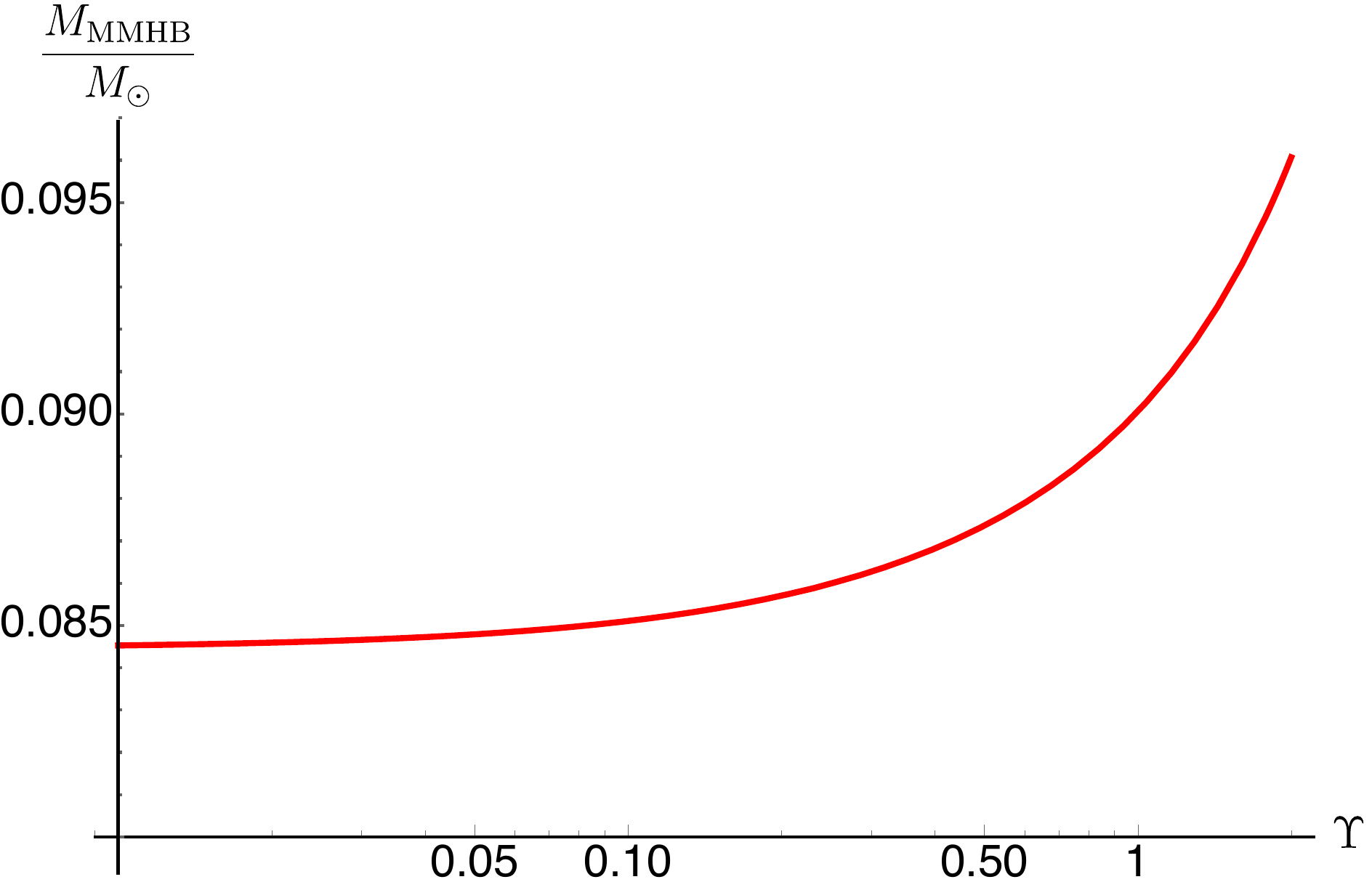}
\caption{The MMHB as a function of $\Upsilon$.}\label{fig:mass}
\end{figure}

The condition for stable hydrogen burning is $\lhb=L\eee$, and so using Eqns. (\ref{eq:LHB}) and (\ref{eq:L}) we have
\begin{equation}\label{eq:etafunc}
3.76 M_{-1}
=\left[\frac{\left(1+\frac{\Upsilon}{2}\right)}{\kar}\right]^{0.11}\left(1+\frac{3\Upsilon}{2} 
\right)^{0.14}\frac{\gamma^{1.32}\omega^{0.09}}{\delta^{0.51}}I(\eta), 
\end{equation}
with
\begin{equation}
I(\eta)\equiv \frac{(\alpha+\eta)^{1.509}}{\eta^{1.325}}.
\end{equation}
From here on we set $\kar=1$, which is typical for high-mass brown dwarfs. The stellar composition does not vary between different 
stars by large amounts, and, since equation (\ref{eq:etafunc}) is only a weak function of the opacity, deviations from this value are highly 
sub-dominant to modified gravity effects. Importantly, the function $I(\eta)$ has a unique minimum value of $2.34$ when $\eta=34.7$. This means that 
if $M_{-1}$ is too low, there is no consistent solution to equation (\ref{eq:etafunc}); The MMHB is the smallest mass for which equation 
(\ref{eq:etafunc}) is satisfied. In GR, one has $\gamma=2.357$, $\delta=5.991$ and 
$\omega=2.714$, which gives $M_{\rm MMHB}^{\rm GR}\approx 0.0845M_\odot$. This is extremely close to the results of detailed numerical simulations 
\cite{1963ApJ...137.1121K}, which predict a value of $0.075M_\odot$. In figure \ref{fig:mass} we plot 
the MMHB predicted by scalar-tensor theories as a function of $\Upsilon$. One can see that theories with increasingly large values of $\Upsilon$ 
predict that the MMHB is larger. This is a consequence of the reduced gravity. When $\Upsilon$ is increased, the central temperature and density at 
fixed mass is reduced since less nuclear burning is needed to provide the pressure gradient to support the star\footnote{Note from equation 
(\ref{eq:L}) that the luminosity of the photosphere is largely independent of the modifications of gravity and scales in the same manner as the 
opacity. Most of the deviations from GR are due to changes in the hydrogen burning rate.}. One then needs a larger mass to achieve the conditions 
necessary for hydrogen burning. 

It is evident that there are large changes in the MMHB even for small deviations from GR. 
% In particular, the MMHB can be larger 
% than $0.01M_\odot$ in alternative theories with $\Upsilon\gsim 0.15$. This is in clear tension with observations. 
There have been several low-mass Red Dwarf stars observed in our local neighbourhood with masses in the range 
$0.08$--$0.2M_\odot$ (see 
\cite{Coppenbarger:1994us,1996A&A...315..418B,Segransan:2000jq,Delfosse:2000jr,Martinache:2006ud,hamilton2012our,2012ApJ...747..144M} and 
references therein). The lowest mass observed M-dwarf is Gl 866 C \cite{Segransan:2000jq}, which has a mass of 
$0.0930\pm0.0008\,M_\odot$\footnote{There are some stars with measured mass $M\sim0.08M_\odot$, but the upper limit given by 
experimental errors exceeds the upper limit for this object.}. Using equation (\ref{eq:etafunc}), one finds that this value is achieved when 
$\Upsilon\approx1.6$, and so values larger than this are excluded.  

% % and so one can exclude the entire class of models with $\Upsilon>0.15$. This is the first model-independent constraint on the 
% % most general scalar-tensor theories of gravity.
% 
As with any astrophysical test of gravity, one must be mindful of possible degeneracies that can potentially mimic the novel effects, or act to 
negate 
them. Here, there are few. Changing $\Upsilon$ is partly degenerate with increasing the opacity but the MMHB is only weakly sensitive to this: it 
scales like $\kappa^{0.11}$, and cannot negate the changes coming from the effects of modified gravity on the structure of the star. The MMHB is 
weakly 
dependent on the amount of stellar rotation, which is absent in our model. Rotation acts to increase the 
MMHB \cite{1992ApJ...393..258S,1970A&A.....8...50K}, and so cannot act to negate the effects of modified gravity. Therefore, it is not a 
caveat to the MMHB predicted in alternative gravity theories. 

Finally, one may worry that the empirical mass determination of low mass objects 
intrinsically assumes GR. This is not the case. The general theories considered here only exhibit deviations from GR inside astrophysical 
bodies. Outside, Newtonian physics applies. Many of the stars referenced above exist either in eclipsing binary systems or have smaller satellites. 
In each case, Newtonian mechanics is used to measure their masses and not their intrinsic properties. In other cases, photometry is used to measure 
the mass using the Mass-Luminosity relation. Examination of Eqn. (\ref{eq:L}) reveals that this does depend on the theory of gravity, however, the 
relation used to calculate the photometric mass is an empirical fit to observations using stars of known mass found using the eclipsing 
binary technique \cite{1993AJ....106..773H}. As such, masses found using this technique are independent of the theory of gravity. 

In summary, we have shown here that the most general scalar-tensor theories of gravity predict that the onset of hydrogen burning in low mass stars 
occurs at higher masses than general relativity predicts. Red Dwarf stars with masses larger than $0.08 M_\odot$ have been observed, and, indeed, this 
value is compatible with general relativity. The absence of 
large degeneracies with non-gravitational physics allows us to confidently rule out the entire region of parameter space where 
$\Upsilon\gsim1.6$. This has two important implications. First, it rules out the possibility that scalar-tensor theories can alter the properties 
of main- and post-main sequence stars because deviations from GR are negligible when $\Upsilon\lsim \mathcal{O}(1)$ \cite{Koyama:2015oma}. Similarly, there are 
only 
negligible effects on the rotation curves of galaxies and the lensing of light by dark matter haloes. 

Second, there are cosmological implications. As remarked above, $\Upsilon$ is given by a specific combination of the parameters appearing in the 
EFT of dark energy. These parameters completely characterise the cosmology on linear scales, and are the focus of upcoming surveys aimed at testing 
gravity in this regime. Using the novel effect presented here, we can place the independent constraint 
\begin{equation}
 \frac{\alpha_H^2}{\alpha_H-\alpha_T-\alpha_B(1+\alpha_T)}\lsim 0.4.
\end{equation}
This directly restricts the possible deviations from general relativity on cosmological scales, and has the potential to rule out competitors to 
$\Lambda$CDM. Indeed, one example of a commonly studied alternative to $\Lambda$CDM is the covariant quartic galileon, which admits a 
self-accelerating de-Sitter solution with $\Upsilon=1/3$. The constraint we have placed here is an upper bound on $\Upsilon$. It is interesting to 
note that stable stellar configurations can not exist when $\Upsilon\le -2/3$ \cite{Saito:2015fza}, and so only a narrow 
window for $\Upsilon$ remains.

% The constraint obtained here is strong enough to completely rule this model out.
% For completeness, we note that there is a small region of parameter space where $\Upsilon<0$ ($\Upsilon$ is bounded from below by $\Upsilon\ge-2/3$ 
% \cite{Saito:2015fza}). Theories where $\alpha_H-\alpha_T-\alpha_B(1+\alpha_T)<0$ are not constrained by the MMHB since they predict a lower MMHB than 
% GR, and are hence compatible with the lowest mass hydrogen burning objects observed. 
% }

% As 
% remarked above, $\Upsilon$ is related to the new mass-scale appearing in the action\footnote{The strong-coupling scale $\Lambda$.} and the 
% cosmological time-derivative of the scalar. This is precisely the combination that is constrained using cosmological probes of gravity 
% (see \cite{Barreira:2013jma} for example). In particular, the best-fitting cosmological model predicts a value for this 
% quantity that is independent of the constraint obtained here. 

What is perhaps most interesting is that the behaviour of local objects have the power to elucidate the nature of dark energy. This remains elusive, 
and the constraint we have obtained here clearly has the potential to drastically reduce the space of viable alternative gravity theories.

\begin{acknowledgments}
 {\bf Acknowledgments: }I am indebted to Olivier Asin for pointing out a typo in equation \eqref{eq:MLE} which led to figure \ref{fig:mass} as well as the bound on $\Upsilon$ being incorrect in the published version of this work. I would like to thank Kazuya Koyama for a careful reading of the paper, and for several enlightening discussions. I am grateful to Vincenzo Salzano for spotting several typos.
\end{acknowledgments}

\bibliography{ref}

\end{document}